\documentclass[aoas]{imsart}

\RequirePackage{amsthm,amsmath,amsfonts,amssymb}
\RequirePackage[authoryear]{natbib}
 \RequirePackage{booktabs}
 \RequirePackage{algorithm}
 \usepackage{algpseudocode} 
\usepackage{multirow} 
\usepackage{tabularray}
 \usepackage{graphicx}   % for \includegraphics
\usepackage{subcaption} % modern subfigure support
\RequirePackage{amsmath}
\usepackage{bbm}

\startlocaldefs
\endlocaldefs

\begin{document}

\begin{frontmatter}
\title{An Integrative Approach for Subtyping Mental Disorders Using Multimodal Data}
%\title{A sample article title with some additional note\thanksref{T1}}
\runtitle{MINDS: Mixed-Type Bayesian Subtyping Framework}
%\thankstext{T1}{A sample of additional note to the title.}

\begin{aug}
%%%%%%%%%%%%%%%%%%%%%%%%%%%%%%%%%%%%%%%%%%%%%%%
%% Only one address is permitted per author. %%
%% Only division, organization and e-mail is %%
%% included in the address.                  %%
%% Additional information such as            %%
%% identifying the corresponding author must %%
%% be included in in the Acknowledgments     %%
%% section if necessary.                     %%

\author[A]{\fnms{Yinjun}~\snm{Zhao} \ead[label=e1]{yz3503@cumc.columbia.edu}},
\author[A]{\fnms{Yuanjia}~\snm{Wang}\ead[label=e2]{yw2016@cumc.columbia.edu}\orcid{0000-0000-0000-0000}}
\and
\author[B]{\fnms{Ying}~\snm{Liu}\thanks{[\textbf{Corresponding author indication should be put in the Acknowledgment section if necessary.}]}\ead[label=e3]{summeryingl@gmail.com}}
%%%%%%%%%%%%%%%%%%%%%%%%%%%%%%%%%%%%%%%%%%%%%%
%% Addresses                                %%
%%%%%%%%%%%%%%%%%%%%%%%%%%%%%%%%%%%%%%%%%%%%%%
\address[A]{Department of Biostatistics, Columbia University\printead[presep={ ,\ }]{e1}}

\address[B]{Statistical Innovation Group AbbVie Inc. \printead[presep={,\ }]{e2,e3}}
\end{aug}

\begin{abstract}
Understanding the biological and behavioral heterogeneity underlying psychiatric disorders is critical for advancing precision diagnosis, treatment, and prevention. This paper addresses the scientific question of how multimodal data—spanning clinical, cognitive, and neuroimaging measures—can be integrated to identify biologically meaningful subtypes of mental disorders. We introduce Mixed INtegrative Data Subtyping (MINDS), a Bayesian hierarchical model designed to jointly analyze mixed-type data for simultaneous dimension reduction and clustering. Using data from the Adolescent Brain Cognitive Development (ABCD) Study, MINDS integrates clinical symptoms, cognitive performance, and brain structure measures to subtype Attention-Deficit/Hyperactivity Disorder (ADHD) and Obsessive-Compulsive Disorder (OCD). Our method leverages P\'olya–Gamma augmentation for computational efficiency and robust inference. Simulations demonstrate improved stability and accuracy compared to existing clustering approaches. Application to the ABCD data reveals clinically interpretable subtypes of ADHD and OCD with distinct cognitive and neurodevelopmental profiles. These findings show how integrative multimodal modeling can enhance the reproducibility and clinical relevance of psychiatric subtyping, supporting data-driven policies for early identification and targeted interventions in mental health.
\end{abstract}


\end{frontmatter}
%%%%%%%%%%%%%%%%%%%%%%%%%%%%%%%%%%%%%%%%%%%%%%
%% Please use \tableofcontents for articles %%
%% with 50 pages and more                   %%
%%%%%%%%%%%%%%%%%%%%%%%%%%%%%%%%%%%%%%%%%%%%%%
%\tableofcontents

%%%%%%%%%%%%%%%%%%%%%%%%%%%%%%%%%%%%%%%%%%%%%%
%%%% Main text entry area:
\section{Introduction}
\label{s:intro}

Understanding why individuals with the same psychiatric diagnosis often present distinct symptoms, neurobiological profiles, and treatment responses remains a major challenge in mental health research \citep{schnack2019improving, marquand2016understanding, review, kozak2016nimh}. This paper addresses the scientific question of how multimodal information, clinical, cognitive, and neuroimaging data, can be integrated to identify reproducible subtypes of mental disorders. Such subtyping is essential for advancing precision psychiatry, which aims to move beyond one-size-fits-all diagnostic categories toward individualized approaches to prevention and treatment.

The importance of this problem is widely recognized. Psychiatric disorders such as Attention-Deficit/Hyperactivity Disorder (ADHD) and Obsessive-Compulsive Disorder (OCD) are highly heterogeneous, and current diagnostic systems based solely on symptoms fail to capture their biological diversity \citep{kozak2016nimh}. This heterogeneity contributes to inconsistent treatment responses and hinders the development of effective, targeted interventions. The Research Domain Criteria (RDoC) framework \citep{Insel_and_Cuthbert}, launched by the U.S. National Institute of Mental Health, emphasizes a dimensional, data-driven approach, integrating behavioral, cognitive, and neurobiological measures to uncover the mechanisms underlying mental illness and improve diagnostic precision. 
As the largest ongoing longitudinal study of brain development and child health in the United States, initiated in 2015 and supported by the National Institutes of Health, the Adolescent Brain Cognitive Development (ABCD) Study follows nearly 12,000 children recruited at ages 9–10 from 21 research sites nationwide \citep{casey2018adolescent}. Each participant undergoes repeated assessments encompassing clinical and behavioral interviews, cognitive testing, multimodal neuroimaging, biospecimen collection, and environmental and demographic surveys. This rich, multimodal design provides an unprecedented opportunity to examine how neurobiological, cognitive, and environmental factors jointly shape mental health trajectories during adolescence. 

Despite these opportunities, existing clustering methods for psychiatric data face critical limitations.
Conventional distance-based subtyping approaches can be extended to handle mixed-type data by using distinct similarity measures for each domain, for example, Euclidean distance for continuous variables and Dice or Jaccard coefficients for binary variables.
A well-known metric in this class is Gower’s distance, which computes overall similarity across heterogeneous data types and is widely used for mixed-data clustering (Gower, 1971).
Another common strategy is a two-step feature-based approach.
In the first step, features are transformed or reduced in dimension using techniques such as:
(1) binary-to-continuous transformation (e.g., logistic regression to estimate latent liabilities \citep{mclachlan2000finite});
(2) factor analysis with polychoric or polyserial correlations; or
(3) principal component analysis (PCA) based on Gower’s distance or Joint and Individual Variation Explained (JIVE) for multi-block data integration \citep{lock2013joint}.
This yields a compact, continuous latent representation of mixed-type data.
In the second step, a conventional clustering algorithm (e.g., K-means or hierarchical clustering) is applied to the transformed features.

Recent model-based frameworks such as iClusterBayes \citep{mo2018fully} also follow this two-step paradigm.
They jointly model continuous and binary variables through latent factor modeling to create a shared low-dimensional representation, followed by K-means clustering within that latent space.
While such approaches effectively leverage domain-specific data structures, clustering on intermediate features rather than the original joint likelihood can discard shared information across modalities, leading to suboptimal efficiency and reproducibility.
Alternatively, latent class analysis (LCA) provides a formal probabilistic framework for mixed-type data by modeling both continuous and categorical variables jointly.
However, fitting LCA models can be computationally prohibitive because the joint likelihood often involving multinomial and Gaussian components lacks a closed-form expression.
Numerical integration is therefore required for parameter estimation, and this becomes infeasible when the latent space is high-dimensional \citep{hagenaars2002applied}.
These limitations highlight the need for a unified, computationally efficient framework that can directly model mixed-type data, retain shared cross-domain structure, and provide stable, interpretable clustering.

 To address these gaps, we propose a new Bayesian hierarchical framework Mixed INtegrative Data Subtyping (MINDS) that performs joint modeling, dimension reduction, and clustering for mixed-type data. Specifically, MINDS integrates binary clinical items (e.g., KSADS psychiatric symptoms), continuous cognitive measures (e.g., NIH Toolbox tasks), and cortical thickness features from neuroimaging, enabling a coherent probabilistic treatment of multimodal inputs. The model employs Pólya–Gamma data augmentation \citep{polya_gamma} to facilitate efficient Gibbs sampling and posterior inference. Through extensive simulation studies, we show that MINDS achieves greater accuracy and stability than competing methods. In this work, we leverage the ABCD dataset as an ideal testbed for the proposed framework, as it combines heterogeneous data types, binary clinical indicators, continuous cognitive scores, and neuroimaging features within a single, harmonized cohort suitable for integrative statistical modeling. Applied to the dataset, MINDS identifies biologically and behaviorally meaningful subtypes of ADHD and OCD with distinct neurocognitive profiles, offering new insights into mechanisms of comorbidity and heterogeneity.

In summary, this study contributes both methodologically by providing a general statistical framework for integrative clustering of mixed data and substantively by revealing clinically relevant subtypes that can guide future research and policy efforts in precision psychiatry. 
The rest of this paper is organized as follows. Section \ref{s:model} introduces the proposed models and Pólya-Gamma augmentation for integrating categorical/binary modality, as well as the details of the algorithm. Section \ref{s:simulation} presents simulation studies conducted to assess the consistency of the estimator and the efficiency of our method. Here, we also compare the performance of our approach to several alternative methods. In Section \ref{s:application}, we apply our methods to the ABCD study, subtyping adolescent participants based on symptoms, neuro-cognitive measures, and brain measures related to ADHD and OCD. We conclude with limitations and extensions in Section 5. 

\section{Model}
\label{s:model}
\subsection{The IRT model}
%{\color{red} Need to introduce the relationship of IRT with our method and why IRT in relevant here in our motivating use case. }

The Item Response Theory \citep{irt} is a framework used to model the relationship between latent traits, i.e., unobserved characteristics or attributes, such as attention or cognitive control, and their manifestations in observed variables, such as responses to items in questionnaires or clinical instruments. The IRT is widely used in educational testing, psychology, psychometrics, and other fields where assessments and questionnaires are utilized to measure underlying traits, including ability, attitude, or personality \citep{van1997item}. It is a suitable model to capture the relationship between binary responses in the KSADS to identify latent traits of attention and cognitive control. MINDS addresses multimodal integration by extending the IRT with Bayesian hierarchical joint modeling for clustering and integrating different data types. In the following, we briefly introduce commonly used IRT models and how they are adapted in MINDS. 
 
In the Rasch model, for a binary item, the probability that a person $i$ with a latent trait $\theta_i$ (e.g., attention) endorsing item $j$ with difficulty parameter $\delta_j$ is given by a logistic regression with random effects, 
\begin{equation}
P(Y_{ij}=1|\theta_i, \delta_j) = \frac{e^{\theta_i-\delta_j}}{1+e^{\theta_i-\delta_j}}.
\label{eq17}
\end{equation}
The relationship specified in  Equation \eqref{eq17} is also called the item response function (IRF). 
To accommodate the non-unity slope of the IRF, an additional parameter $\alpha$ is introduced to generalize the Rasch model to a one-parameter logistic (1PL) model such that
\begin{equation*}
P(Y_{ij}=1|\theta_i, \alpha,\delta_j) = \frac{e^{\alpha(\theta_i-\delta_j)}}{1+e^{\alpha(\theta_i-\delta_j)}}
\label{eq18}.
\end{equation*}
When $\alpha$ varies across items, the 1PL model becomes a two-parameter logistic (2PL) model as
\begin{equation*}
P(Y_{ij}=1|\theta_i, \alpha_j,\delta_j) = \frac{e^{\alpha_j(\theta_i-\delta_j)}}{1+e^{\alpha_j(\theta_i-\delta_j)}}
\label{eq19}.
\end{equation*}
An item's $\alpha_j$ characterizes how well the item can differentiate among individuals located at different points on the continuum. Thus, $\alpha_j$ is also referred to as the item $j$’s discrimination parameter.

The multidimensional IRT (MIRT) model extends the traditional unidimensional IRT to account for multiple latent traits, such as attention and cognitive control, as
\begin{equation*}
P(Y_{ij}=1|\boldsymbol{\theta}_i, \boldsymbol{\alpha}_j,\gamma_j) = \frac{e^{\sum_{l=1}^L \alpha_{jl}\theta_{il}+\gamma_j}}{1+e^{\sum_{l=1}^L \alpha_{jl}\theta_{il}+\gamma_j}}
= \frac{e^{\boldsymbol{\alpha}_j'\boldsymbol{\theta}_i+\gamma_j}}{1+e^{\boldsymbol{\alpha}_j'\boldsymbol{\theta}_i+\gamma_j}},
\end{equation*}
where $\boldsymbol{\theta}_i=\{\theta_{il}\}_{l=1,\cdots,L}$ is the 
$L$-dimensional vector of latent traits for person $i$, $\boldsymbol{\alpha}_j=\{\alpha_{jl}\}_{l=1,\cdots,L}$ is the 
$L$-dimensional vector of discrimination parameters for item $j$, and $\gamma_j=-\sum_{l=1}^L \alpha_{jl}\delta_{jl}$ is the difficulty parameter for item $j$, also referred to as the item $j$'s threshold. This is useful when test items are conceptualized to assess multiple underlying abilities or attributes.

\subsection{Latent mixture model with single modality}\label{sec:basic}
%{\color{red}need a transition sentence here from the previous IRT model} 
The conventional MIRT models can only conduct dimension reduction for categorical data and do not accommodate mixture distributions. Here, we extend them to incorporate continuous variables and conduct clustering simultaneously.
%Traditional IRT models are limited to dimension reduction for categorical data and do not support clustering. In this work, we extend IRT models to enable both dimension reduction and clustering simultaneously.
We first describe the modeling for binary, ordinal, or continuous measures, respectively, and then introduce how to integrate different data types.

Let $Y_{ij}$ denote the binary modality's $j$th item in a clinical instrument from the $i$th subject, where $i=1,\cdots, N_b$, and $j=1, \cdots, N_d$.  We propose a latent mixture model inspired by MIRT for the binary modality as 
\begin{eqnarray}
Y_{ij}| X, a_{j},Z_i, b_i, V_{.j}\sim 
\text{Bernoulli} \left( \frac {e^{ (Z_i^T X+ b^T_i)V_{.j}-a_{j}}} {1+ e^{(Z_i^T X +b^T_i) V_{.j} -a_{j}}} \right), \label{m_binary}
\end{eqnarray}
where $X$ is a $N_c\times N_t$ matrix representing the  $N_c$ cluster centers, each of which is a $N_t$-dimensional vector. Here, $Z_i$ is an indicator of the $i$th subject's unobserved cluster membership assumed to follow a categorical distribution, $p(Z_i=e_k)=\theta_k$, 
where $e_k$ is the $k$-th standard basis vector in $\mathbb{R}^K, i.e., 
e_k = (0, 0, \dots, 1, \dots, 0)^\top$ with the 1 in the $k$-th position. $b_i$ is the $i$th subject's latent construct, e.g., cognitive control ability that deviates from his or her cluster mean ability. Furthermore, $V_{.j}$ is an $N_t$-dimensional loading vector on latent construct for the $j$th item representing how well the items can differentiate among individuals at different points of latent constructs, and $a_j$ represents the $j$th item's threshold. Note that the above model reduces to the MIRT model when there is a single cluster.   
A similar method under a cumulative model can be developed for ordinal measures as
\begin{eqnarray*}
P(Y_{ij}\leq \gamma_t| X, a_{jt},Z_i, b_i, V_{.j} )=
  \frac {e^{ (Z_i^T X+ b^T_i)V_{.j}-a_{jt}}} {1+ e^{(Z_i^T X +b^T_i) V_{.j} -a_{jt}}}, 
\end{eqnarray*}
where $\gamma_t=\sum_{s=1}^{t} \pi_s, \quad t=1, \cdots, T_j$, and  $\pi_t$  represents the probability that  $Y_{ij}$  belongs to $t$th category,  and $a_{jt}$ is the threshold parameter for $j$th item belonging to $t$th category.
For modeling $N_d$-dimensional continuous measures $Y_{ij}$, we propose a multivariate Gaussian model as
$$
Y_{ij}| X, a_{j},Z_i, b_i, U_{.j}\sim 
\text{MVN} (( Z_i^T X + b^T_i)U_{.j}-a_j, \sigma^2_j).
$$ The interpretations of $a_{j},Z_i, b_i, U_{.j}$ are similar to $X, a_{j},Z_i, b_i, V_{.j}$ in model \eqref{m_binary}. In these models, individuals in different subtypes have distinct mean latent traits (i.e., cluster center $X$) while sharing other parameters such as discriminant and threshold values.

%{\color{red}may need to explain why Polya-Gamma is necessary or more efficient in our algorithm}

Since the binary modality’s distribution does not belong to the Gaussian family, the posterior distribution is not conjugate to the prior, making Gibbs sampling inapplicable.
Pólya-Gamma augmentation has been proposed to yield a Gibbs sampler for the Bayesian logistic model \citep{polya_gamma}. %Before we introduce how the Pólya-Gamma augmentation will be implemented in our model, we introduce its key ideas. 
A random variable $Y$ has a Pólya-Gamma distribution with parameters $b>0$ and $c\in\mathbb{R}$, denoted as $Y\sim \text{PG}(b,c)$, if 
\begin{equation*}
Y \overset{D}{=}  \frac{1}{2\pi^2}\sum_{k=1}^\infty \frac{g_k}{(k-1/2)^2+c^2/(4\pi^2)},
\label{eq4.1}
\end{equation*}
where $g_k\sim \text{Ga}(b,1)$ are independent gamma random variables, and $\overset{D}{=}$ indicates equality in distribution. Let $p(\omega)$ denote the density of the random variable $\omega\sim \text{PG}(b,0)$, $b>0$. Then the following integral identity holds for all $a\in\mathbb{R}$:
\begin{equation}
\frac{(e^\psi)^a}{(1+e^\psi)^b} = 2^{-b} e^{\kappa \psi}\int_0^\infty e^{-\omega \psi^2/2}\, p(\omega)\, d\omega
,\label{eq4.2}
\end{equation}
where $\kappa=a-b/2$.
Moreover, the conditional distribution
$p(\omega|\psi)$, where $\psi$ is the exponential part in \eqref{eq4.2}, is also in the Pólya-Gamma class, i.e., \begin{equation}
    (\omega\,|\,\psi)\sim \text{PG}(b,\psi).
    \label{prop1}
\end{equation}
The properties of \eqref{eq4.2} \eqref{prop1} suggest a simple strategy for Gibbs sampling across a wide class of binomial models, that is, Gaussian draws for the main parameters and Pólya-Gamma draws for a single layer of latent variables. 
% {\color{red}
% Specifically, denote $y_i$ be the number of successes, $n_i$ the number of trials, and $x_i=(x_{i1},\cdots, x_{ip})$ the vector of regressors for observation $i\in\{1,\cdots,N\}$. Let $y_i\sim \text{Binom}(n_i, 1/\{1+e^{-\psi_i}\})$, where $\psi_i=x_i^T \beta$ are the log odds of success. Suppose $\beta$ have a Gaussian prior, $\beta\sim \text{N}(b, B)$. To sample from the posterior distribution using the Pólya-Gamma method, we simply iterate two steps:
% \begin{equation*}
% (\omega_i\,|\,\beta) \sim \text{PG}(n_i,\, x_i^T \beta)\,\,\,\,\text{and}\,\,\,\,
% (\beta\,|\,y,\omega) \sim \text{N}(m_\omega,\, V_\omega),
% \label{eq4.4}
% \end{equation*}
% where 
% $V_\omega = (X^T \Omega X + B^{-1})^{-1}$,
% $m_\omega = V_\omega (X^T \kappa + B^{-1}b)$, 
% $\kappa = (y_1-n_1/2, \cdots, y_N-n_N/2)$, and $\Omega$ is the diagonal matrix of $\omega_i$'s.} 
Under model \eqref{m_binary}, the conditional distribution of $Y$ given all the rest of parameters is
\begin{equation*}
    P(Y|X, Z_i, b_i, a_j, V_{.j})=\prod_{i=1,\cdots,N_b}\prod_{j=1,\cdots,N_d}\frac{e^{\psi_{ij}y_{ij}}}{1+e^{\psi_{ij}}},
\end{equation*}
By Equation \eqref{eq4.2}, the conditional distribution can be expressed as 
\begin{equation}
    P(Y|X, Z_i, b_i, a_j, V_{.j})=\frac{1}{2}\prod_i\prod_j \exp\left\{(y_{ij}-1/2)\psi_{ij}\right\}E_{\omega_{ij}\sim PG(1,0)}\exp\left\{-\psi_{ij}^2\omega_{ij}/2\right\}
    \label{py_1},
\end{equation}
where $\psi_{ij} =(Z_i^T X+ b^T_i)V_{.j}-a_{j}$.

Given $\omega_{ij}$, the expectation in Equation \eqref{py_1} is $\exp\left\{-\psi_{ij}^2 \omega_{ij}/2\right\}$, which leads
the conditional probability of $Y_{ij}$ in model \eqref{m_binary} follows the Gaussian distribution.
\begin{equation}
    P(Y|\omega_{ij}, X, Z_i, b_i, a_j, V_{.j})=\frac{1}{2}\prod_i\prod_j \exp\left\{(y_{ij}-1/2)\psi_{ij}\right\}\exp\left\{-\psi_{ij}^2\omega_{ij}/2\right\}
    \label{py_2}.
\end{equation}
By the property of the Pólya-gamma distribution in Equation \eqref{prop1}, we have   
$$
\omega_{ij}|X, Z_i, b_i, a_j, V_{.j}\sim PG(1,\psi_{ij} ).$$ 
In this sense, all parameters' posterior distributions are conjugate to their prior distributions, so the Gibbs sampling can be adopted for parameter estimation.

\subsection{Latent mixture model with multi-modalities}
Model \eqref{m_binary} only handles a single modality. Here, we extend the single-modality model to handle more complex multimodal data using discrete clinical and continuous behavioral measures.  We propose a joint model that integrates binary and continuous modalities as follows
\begin{eqnarray}
Y_{i\cdot}^{(1)}| X,b_{i}, V ,Z_i, a^{(1)}&\sim 
&\text{Bernoulli} \left( \frac {e^{(Z_i^TX+b^T_{i})V-a^{(1)}}} {1+ e^{ (Z_i^TX+b^T_{i})V-a^{(1)}}} \right),\label{eq4.8}\\
Y_{i\cdot}^{(2)}| X, b_{i},U, Z_i, a^{(2)}, \Sigma &\sim& 
\text{Normal} \left((Z_i^TX+b^T_{i})U-a^{(2)}, \Sigma\right),\label{eq4.9}
\end{eqnarray}
where $X$ is a matrix of cluster centers shared by these two modalities, $Z_i$ is subject $i$th's cluster membership, $b_{i}$ is $N_t$-dimensional subject's latent construct, $V$ and $U$ are $N_t\times N_{d_1} $-dimensional and $N_t\times N_{d_2}$-dimensional loading matrix on cluster center $X$ for binary and continuous modality, respectively. $N_{d_1}$ represents the number of items in binary modality, and $N_{d_2}$ represents the number of measures in continuous modality. $a^{(1)}$ and $a^{(2)}$ represent item difficulty for two different modalities. For continuous modality,
 $\Sigma$ is a diagonal covariance matrix with diagonal entries $\sigma_j^2$, $j=1,\cdots, N_{d_2}$.
In this model, we assume the two data-type modalities share the cluster center $X$, and each subject belongs to the same cluster under different modalities. 

%To address this issue, we follow the standard practice in factor analysis to impose constraints on certain item loadings. In the two-dimensional case, setting one item's loading parameter on one axis to zero, i.e., $V_{11}=0$., fixes the orientation of the other axis. More generally, with $N_t$ latent constructs, constraints on $N_t-1$ items are necessary to address the rotational indeterminacy. {\color{red}For example, for the $k$th item, restrict its loadings on the first $N_t-1$ latent constructs to be zero; for another item, restrict its loadings on $N_t-2$ latent constructs to be zero, and so on and so forth. }

%For parameter identifiability, we assume the variance of the subject's specific ability $b_i$ to be known and set $X_{11}$ in cluster center $X$ to be known. 
We use a Bayesian hierarchical algorithm to estimate the parameters. 
We assume the following prior distributions for the parameters in \eqref{eq4.8} and \eqref{eq4.9} as that
$Z_i$ follows categorical distribution with hyperparameter $\theta$; 
$b_{i}$ follows $N(0, \mathcal{I}(\sigma_b^2))$;
$X, V, U$ are from multivariate normal distributions, $N(\mu_X, \mathcal{I}(\sigma_x^2))$, $N(\mu_V, \mathcal{I}(\sigma_v^2))$, and $N(\mu_U, \mathcal{I}(\sigma_u^2))$, respectively; 
$a^{(1)}, a^{(2)}$ are from multivariate normal, $N(0, \mathcal{I}(\sigma_a^2))$, respectively. We assume $\theta$ follows Dirichlet distribution with parameters of equal weight $1/N_c$, and 
$\sigma_b^2$ and $\sigma_j^2$ follow Inverse Gamma distribution $IG(\alpha_b, \beta_b
)$ and $IG(\alpha_j, \beta_j
)$, respectively. 

The conditional distribution of $Y$ given all the rest parameters in the joint model \eqref{eq4.8}-\eqref{eq4.9} is
\begin{eqnarray*}
P(Y|X, Z_i, b_i, a^{(1)}, a^{(2)}, V,U)=\prod_{i=1,\cdots,N_b}\prod_{j=1,\cdots,N_{d_1}} p\left(y_{ij}^{(1)}|rest\right) \cdot \\
\prod_{i=1,\cdots,N_b} \prod_{j=1,\cdots,N_{d_2}}p\left(y_{ij}^{(2)}|rest\right).\label{py_3}
\end{eqnarray*}
A similar method is used as for model \eqref{m_binary}, which introduces a latent variable $\omega$ following the P\'olya-Gamma distribution such that  
$
\omega_{ij}|X, Z_i, b_i, a_j^{(1)}, V_{.j}\sim PG(1,\psi_{ij} )$, where $\psi_{ij} =(Z_i^T X+ b_i^T)V_{.j}-a_{j}^{(1)}$.
Similar to equation \eqref{py_2}, the conditional distribution of $Y_{ij}^{(1)}$ given $\omega$ and all the rest of the parameters in model \eqref{eq4.8} is from a Gaussian family. In this sense, the posterior distributions of all the parameters $X, b_i, a_j^{(1)}, a_j^{(2)}, V_{.j}, U_{.j}, \sigma_j^2$ in the joint model are conjugate to their prior distributions. The conditional probability that the $i$th subject belongs to the $k$th cluster, given the rest of the parameters, denoted as $\pi_{ik}$, is expressed as:
%\begin{eqnarray*}
%\pi_{ik}&=&\prod_{j=1,\cdots,N_{d_1}} \exp\left(-\frac{\omega_{ij}}{2}\left[(X_k+b^T_i) V_{.j}-\frac{\kappa_{ij}}{\omega_{ij}} -a_j^{(1)}\right]^2 \right)\cdot \\
%&&\prod_{j=1,\cdots,N_{d_2}} \frac{1}{\sigma_j}\exp\left(-\frac{1}{2\sigma_j^2}\left[(X_k+b^T_i) U_{.j}-y_{ij}^{(2)} -a_j^{(2)}\right]^2 \right)\cdot P(Z_i=k|\theta)\cdot P(\theta),
%\end{eqnarray*}
\begin{eqnarray*}
\pi_{ik}=&&\prod_{j=1,\cdots,N_{d_1}} \exp\left(-\frac{\omega_{ij}}{2}\left[(X_k+b^T_i) V_{.j}-\frac{\kappa_{ij}}{\omega_{ij}} -a_j^{(1)}\right]^2 \right)\cdot \\
&&\prod_{j=1,\cdots,N_{d_2}} \frac{1}{\sigma_j}\exp\left(-\frac{1}{2\sigma_j^2}\left[(X_k+b^T_i) U_{.j}-y_{ij}^{(2)} -a_j^{(2)}\right]^2 \right)\cdot \\
&&P(Z_i=e_k|\theta)\cdot P(\theta),
\end{eqnarray*}
which indicates that the posterior distribution of $Z_i$ is from Multi$(1, \pi_i)$,$\quad \pi_i=(\pi_{i1}, \cdots, \pi_{iN_c})$.
All the parameters' posterior distributions can be expressed in closed forms so that the Gibbs sampling can be adopted for parameter estimation. Details of the posterior distributions are in section \ref{algorithm}.

\subsection{Algorithm}\label{algorithm}
We derive Gibbs sampling steps for the joint model \eqref{eq4.8}\&\eqref{eq4.9} in Algorithm \ref{alg:gibbs}. This algorithm can be easily modified to accommodate the single modality model \eqref{m_binary}, a simpler version of the joint model. 
%Deviance Information Criterion (DIC) \citep{spiegelhalter2014deviance} is used to determine the optimal number of clusters $N_c$.
Information Criterion (IC), a selection information criterion for Bayesian models \citep{ando2011predictive}, which avoids the problems of over-fitting associated with DIC \citep{spiegelhalter2014deviance}, is used to determine the optimal number of clusters $N_c$.
$$IC =-2E_{\theta|y}[logf(y|\theta)]+2P_D,$$
where $P_D $ is defined as difference between the posterior mean of deviance and the deviance estimated at the posterior mean of the parameters, $2[logf(y|\hat\theta_n))]-2E_{\theta|y}[logf(y|\theta)]$, and $\hat{\theta}_n$ is the posterior mean.
The stationary trace plot is used to examine convergence.

\begin{algorithm}[!t]
\caption{Gibbs sampling steps for joint model}\label{alg:gibbs}
\begin{algorithmic}[1]
\Require Binary items $y^{(1)}$, continuous measures $y^{(2)}$
\Ensure Estimators $\boldsymbol{\lambda}= (X, Z, b, a^{(1)}, a^{(2)}, V, U, \theta, \Sigma)$
\State Initialize $\boldsymbol{\lambda}$
\Repeat
    \State $\omega_{ij} \gets PG\!\left(1,\ (Z_i^\top X+ b_i^\top)V_{.j}-a_{j}^{(1)}\right)$
    \State $X_t \gets N\!\left(V_{X_t}\!\left(\sum_{i=1}^{N_b} Z_iU_t\Sigma \Phi_{i}^{(t)}+\sum_{i=1}^{N_b} Z_iV_t\Omega_i\phi_{i}^{(t)}+B_t^{-1}\mu_{X_t}\right),\ V_{X_t}\right)$\label{post:x}
    \State $Z_i \gets \text{categorical}(\pi_i)$\label{post:z}
    \State $V_{tj} \gets N(B_{tj}Q_{tj},\ B_{tj})$\label{post:v}
    \State $U_{tj} \gets N(D_{tj}R_{tj},\ D_{tj})$\label{post:u}
    \State $b_t \gets N(\mu_{b_t},\ V_{b_t})$\label{post:b}
    \State $a_j^{(1)} \gets N\!\left(V_{a_j^{(1)}} \sum_{i=1}^{N_b} \sigma^{-2}_j r_{ij}^{(1)},\ V_{a_j^{(1)}}\right)$\label{post:a1}
    \State $a_j^{(2)} \gets N\!\left(V_{a_j^{(2)}} \sum_{i=1}^{N_b} \omega_{ij} r_{ij}^{(2)},\ V_{a_j^{(2)}}\right)$\label{post:a2}
    \State $\theta \gets Dir\!\left(\sum_{i=1}^{N_b} \mathbbm{1}_{\{Z_i=e_k\}} + \frac{1}{Nc}\right),\quad k=1,\ldots,Nc$
    \State $\sigma_b^2 \gets IG\!\left(\frac{N_b}{2}+\alpha_b,\ \frac{1}{2}\sum_{i=1}^{N_b} b_i^2+\beta_b\right)$
    \State $M_{ij} \gets y_{ij}^{(2)}-(Z_i^\top X+b_i^\top)U_{.j}-a^{(2)}_j$
    \State $\sigma_j^2 \gets IG\!\left(\frac{N_b}{2}+\alpha_j,\ \frac{1}{2}\sum_{i=1}^{N_b} M_{ij}^2+\beta_j\right)$
\Until{$\boldsymbol{\lambda}$ converges}
\end{algorithmic}
\end{algorithm}
\newpage
In Algorithm \ref{alg:gibbs}, for step \ref{post:x}, 
\begin{eqnarray*}
    \Omega_i&=&\mathcal{I}(\omega_{i1}, \cdots, \omega_{iN_{d_1})},\text{ }\Sigma=\mathcal{I}(\sigma_1^{-2}, \cdots,\sigma_{N_{d_2}}^{-2}),\\
    U_t&=&(U_{t1}, \cdots, U_{tN_{d_2}})^T, \text{ } 
    V_t=(V_{t1}, \cdots, V_{tN_{d_1}})^T, \\
    X_t&=&(X_{1t}, \cdots, X_{N_ct})^T,\\
    \phi_{ij}^{(t)}&=&\frac{\kappa_{ij}}{\omega_{ij}} + a_j^{(1)}- (Z_i^T X\cdot \mathbbm{1}^T_{\{s\neq t\}}+b^T_i)V_{.j}, s=1, \cdots, N_t,\\
    \Phi_{ij}^{(t)}&=&y_{ij}^{(2)} + a_{j}^{(2)}-(Z_i^T X \cdot\mathbbm{1}^T_{\{s\neq t\}}+b^T_i) U_{.j}, s=1, \cdots, N_t, \\
    V_{X_{t}}^{-1}&=& \displaystyle\sum_{i=1}^{N_b}Z_iV_{t}\Omega_iV_{t}^TZ_i^T +B^{-1}_t, \text{ } B_t=\mathcal{I}_{N_c}(\sigma_x^{-2}).
\end{eqnarray*}
For step \ref{post:z}, the conditional probability of $i$th subject belongs to the $k$th cluster center, i.e., $Z_i^T=\mathbbm{1}_{\{s= k\}}, s=1, \cdots, N_c$,  given the rest parameters is 
\begin{eqnarray*}
    \pi_{ik}=P(Z_i=e_k|\theta)P(\theta)=
\prod_{j=1}^{N_{d_1}} \exp\left(-\frac{\omega_{ij}}{2}\left[(X_k+b^T_i) V_{.j}-\frac{\kappa_{ij}}{\omega_{ij}} -a_j^{(1)}\right]^2 \right)\cdot \\
\prod_{j=1}^{N_{d_2}} \frac{1}{\sigma_j}\exp\left(-\frac{1}{2\sigma_j^2}\left[(X_k+b^T_i) U_{.j}-y_{ij}^{(2)} -a_j^{(2)}\right]^2 \right)P(\theta).
\end{eqnarray*}
For steps \ref{post:v} and \ref{post:u},
\begin{eqnarray*}
B_{tj}^{-1}&=&\sigma_v^{-2}+\sum_{i=1}^{N_b} \omega_{ij}A_{it}^2, \text{ } Q_{tj}=\frac{\mu_v}{\sigma_v^2}+ \sum_{i=1}^{N_b}\omega_{ij}A_{it}\psi_{ij}\\
D_{tj}^{-1}&=&\sigma_u^{-2}+\sum_{i=1}^{N_b} \sigma_j^{-2}A_{it}^2, \text{ }
    R_{tj}=\frac{\mu_u}{\sigma_u^{2}}+ \sum_{i=1}^{N_b} \sigma_j^{-2}A_{it}\Psi_{ij},\\ \text{where }\\
    A_{it}&=&Z_i^TX_{t}+b_{it},\\       
    \psi_{ij}^{(t)}&=&\frac{\kappa_{ij}}{\omega_{ij}} + a_j^{(1)}- (Z_i^T X+b^T_i) \cdot\mathbbm{1}^T_{\{s\neq t\}}V_{.j}, s=1, \cdots, N_t,\\
    \Psi_{ij}^{(t)}&=&y_{ij}^{(2)} + a_{j}^{(2)}-(Z_i^T X+b^T_i)\cdot \mathbbm{1}^T_{\{s\neq t\}}U_{.j}, s=1, \cdots, N_t.
\end{eqnarray*}
For step \ref{post:b},
\begin{eqnarray*}
\mu_{b_t}&=&\left\{ V_{b_{it}}\left(\sum_{j=1}^{N_{d_1}}V_{tj}C_{ij}^{(t)}\omega_{ij}+
\sum_{j=1}^{N_{d_2}}U_{tj}D_{ij}^{(t)}\sigma_{j}^{-2}\right)\right\}_{i=1, \cdots, N_b},\\
 V_{b_t}^{-1}&=&\mathcal{I}\{\sigma_{b_t}^{-2}+ \sum_{j=1}^{N_{d_1}}V_{tj}^2\omega_{ij}+
\sum_{j=1}^{N_{d_2}}U_{tj}^2\sigma_{j}^{-2}\},\\
C_{ij}^{(t)}&=& \frac{\kappa_{ij}}{\omega_{ij}}+a_j^{(1)}-(Z_i^TX +b^T_i\cdot\mathbbm{1}^T_{\{s\neq t\}})V_{.j}, \\
D_{ij}^{(t)}&= &y_{ij}^{(2)}+a_j^{(2)}-(Z_i^TX +b^T_i\cdot\mathbbm{1}^T_{\{s\neq t\}})U_{.j},s=1, \cdots, N_t.     
\end{eqnarray*}
For steps \ref{post:a1} and \ref{post:a2}, 
\begin{eqnarray*}
r_{ij}^{(1)}= (Z_i^T X+b_i^T)V_{.j}-\frac{\kappa_{ij}}{\omega_{ij}}, \text{  }
V^{-1}_{a_j^{(1)}}=\sum_{i=1}^{N_b}\omega_{ij} +\sigma_a^{-2},\\
r_{ij}^{(2)}= (Z_i^TX+b_i^T)U_{.j}-y_{ij}^{(2)},\text{  } V_{a_j^{(2)}}^{-1}=\sigma_a^{-2}+N_b\sigma_j^{-2}.
\end{eqnarray*}
This model has metric indeterminacy in the sense that we do not have a metric for $V$ and $U$ with an intrinsic origin or unit.
In addition, the model has a rotational indeterminacy in the sense that the axes of $X$ that represent latent variables are free to rotate about their origin because there is no external reference to fix their orientation. To address these issues and ensure identifiability, we set $X_{11}$ to be one and $a_1^{(Nd_1)}$to be zero and estimate $X$, $U$ and $V$ by solving equation system that 
$$ X_k V-a^{(1)}=\hat{X}*\hat{V}-\hat{a^{(1)}}, X_k U-a^{(2)}=\hat{X}*\hat{U}-\hat{a^{(2)}},$$
where $k=1\cdots N_c$, $\hat{X}, \hat{U}, \hat{V}, \hat{a^{(1)}},  \hat{a^{(2)}}$ are estimation from algorithm \ref{alg:gibbs}. The equation system has unique solution when the number of parameters, $N_c*N_t+N_t*(N_{d_1}+N_{d_2}) + N_{d_1}+N_{d_2} -2$ is not greater than the number of equations, $N_c*(N_{d_1}+N_{d_2})$, i.e. $N_c*N_t+2\leq (N_c-N_t-1)*(N_{d_1}+N_{d_2}))$.  

\section{Simulation studies}
\label{s:simulation}
We conducted extensive simulation studies to evaluate the performance of our proposed algorithm for the joint model. This involved examining the consistency of latent variable estimates, calculating training error on the training data and test errors on testing data when the cluster membership is known, and comparisons with alternative methods.

%We included three groups of settings with a combination of different numbers of binary items and continuous measures: (10, 10), (20, 10), (30, 10), where the first element represents the number of binary items and the second element represents the number of continuous measures. For example, (10, 20) represents $N_{d_1}=10$,  $N_{d_2}=20$. 
We generated three simulation settings, each setting composed of 10 continuous measures and 10, 20, 30 binary items, respectively. 
Three latent constructs were considered for the cluster center $X$ and the $i$th subject's latent construct $b_i$, i.e., $N_t=3$. The elements of the true cluster center $X$ and subject-specific ability $b_i$  were randomly drawn from a uniform distribution $U(0, 2)$ and a Gaussian distribution $N(0, 0.2)$, respectively. Subjects were generated from five clusters, i.e., $N_c=5$, with cluster membership weights $\theta$ to be (0.3, 0.15, 0.15, 0.2, 0.2), based on which we generated true membership $Z_i$ from Categorical$(\theta)$ for subject $i$. The item difficulty parameters $a^{(1)}_{j}$ for binary items and $a^{(2)}_j$ for continuous measures were randomly drawn from U(-0.5, 0.5) and U(-5, 5), respectively. All elements of the loading matrix $V$ for binary items and $U$ for continuous measures were independently randomly drawn from U(0, 2). 
All the prior distributions were non-informative, and all the estimators' initial values were random. 

We conducted 200 replications for each setting and sample size (1,000 and 2,000) to evaluate the consistency of latent variable estimates and training error.
As shown in Web Figures 1-4, the medians of the estimated values for cluster center $X$, loading matrix $U$ and $V$, and item difficulty $a^{(1)}$ and $a^{(2)}$ are close to their true values except for $U_{13}, V_{15}$. As the sample size increases from 1000 to 2000, the variance of the estimators decreases. We compared the root mean squared errors (RMSE) and bias of all parameters for sample size 1,000 and 
2,000. As shown in 
Table \ref{table:rmse&bias}, overall parameters' RMSEs decrease as the sample size increases from 1,000 to 2,000. The other two settings have similar performance, and results are omitted. 

Corresponding to the sample size range of the motivating ABCD study, 10,000 subjects were generated as the test data to assess the testing error by using the same true values of parameters $X$, $U$, $V$, $a^{(1)}$, and $a^{(2)}$ as those used in the generation of the training data. For the subject-specific parameters $Z_i$ and $b_i$, the same distributions with the same parameters $\theta$ and $\sigma_b^2$ as in the training data were used in the generation of the testing data, respectively. To get the predicted cluster membership, we plugged the estimated values of $X$, $U$, $V$, $a^{(1)}$, and $a^{(2)}$ from training data, and iterated the remaining unknown parameters $\omega, Z$, and $b_i$ until convergence. The estimated $Z$ returns the predicted cluster membership. The test errors were calculated using three metrics to evaluate the performance of the clustering.
The three metrics are (a) Bayes error rate, which is defined as the expectation of the event that the predicted category (the one having maximum membership weight) is not equal to the true category; (b) Classification error, the proportion that the predicted category is not equal to the true category; (c) Jaccard distance, defined as the one minus the size of the intersection divided by the size of the union of two label sets. 
For all three metrics, smaller scores imply better performance. 

We compared the results of the MINDS method with six other alternative approaches: iClusterBayes, K-means clustering, Hierarchical clustering (shortened for Hclust), two-step JIVE, two-step K-means clustering, and two-step Hclust. In two-step JIVE, principal component (PC) scores were obtained by JIVE, and then K-means was used to cluster the PC scores. In the two-step K-means, we first conducted factor analysis based on the polychoric correlation, and in the second step, used K-means to cluster the factor scores into $N_c$ clusters. The two-step Hclust differs from the two-step K-means in that, in the second step, it uses Hclust to cluster the factor scores. As shown in Table \ref{table:training_error} \ref{table:testing_error} and Web Table 1, the training and testing errors of MINDS are lower than all the other six methods in all three metrics, which shows that the performance of MINDS is satisfactory. 
\begin{table}[!t]
\centering
\small
\caption{Simulation results from a joint model based on 200 replicates; RMSE and bias are calculated for 1,000 vs. 2,000 training sample size; each model integrated one binary and one continuous modality, incorporating 10–30 binary items and 10 continuous measures.}
\label{table:rmse&bias}
% \usepackage{multirow}
\begin{tabular}{lccccccc}
\toprule
                      & Training Size & Items & $x$     & $V$     & $U$     & $a$     & $\theta$ \\ \midrule
\multirow{6}{*}{BIAS} & 1000          & 10    & 0.054 & 0.046 & 0.024 & 0.273 & 0.006 \\ 
                      & 2000          & 10    & 0.059 & 0.036 & 0.024 & 0.271 & 0.003 \\ \cmidrule{2-8} 
                      & 1000          & 20    & 0.093 & 0.055 & 0.032 & 0.252 & 0.012 \\ 
                      & 2000          & 20    & 0.084 & 0.042 & 0.026 & 0.244 & 0.009 \\ \cmidrule{2-8} 
                      & 1000          & 30    & 0.076 & 0.059 & 0.062 & 0.194 & 0.01  \\ 
                      & 2000          & 30    & 0.056 & 0.042 & 0.034 & 0.187 & 0.001 \\ \midrule
\multirow{6}{*}{RMSE} & 1000          & 10    & 0.158 & 0.156 & 0.09  & 0.314 & 0.008 \\ 
                      & 2000          & 10    & 0.134 & 0.117 & 0.074 & 0.303 & 0.005 \\ \cmidrule{2-8} 
                      & 1000          & 20    & 2.087 & 0.647 & 0.691 & 1.457 & 0.836 \\ 
                      & 2000          & 20    & 2.092 & 0.651 & 0.688 & 1.458 & 0.836 \\ \cmidrule{2-8} 
                      & 1000          & 30    & 2.17  & 0.808 & 0.578 & 1.239 & 0.853 \\ 
                      & 2000          & 30    & 2.161 & 0.808 & 0.571 & 1.243 & 0.863 \\ \bottomrule
\end{tabular}
\end{table}

\begin{table}
\centering
\caption{Mean(sd) of training set Bayes error from MINDS and alternative methods, evaluated across 200 replicates. Training sample sizes ranged from 1,000 to 2,000 subjects. Each model integrated one binary and one continuous modality, incorporating 10–30 binary items and 10 continuous measures.}
\label{table:training_error}
\resizebox{\textwidth}{!}{

\begin{tabular}{lclllllll}
\toprule
\multicolumn{1}{c}{\multirow{2}{*}{\textbf{Items}}} & \multirow{2}{*}{\textbf{Training Size}} & \multicolumn{7}{c}{\textbf{Training Set Bayes Error}} \\  \cmidrule{3-9} 
\multicolumn{1}{c}{} &      & MINDS        & iClusterBayes & Two-step JIVE & K-means      & Hclust       & Two-step K-means & Two-step Hclust \\ \midrule
10                   & 1000 & 0.049(0.03)  & 0.107(0.01)   & 0.135(0.009)  & 0.111(0.044) & 0.152(0.01)  & 0.107(0.042)     & 0.134(0.032)    \\
10                   & 2000 & 0.046(0.024) & 0.104(0.008)  & 0.141(0.006)  & 0.127(0.038) & 0.15(0.009)  & 0.12(0.043)      & 0.145(0.031)    \\ 
20                   & 1000 & 0.028(0.021) & 0.081(0.015)  & 0.168(0.026)  & 0.084(0.062) & 0.144(0.034) & 0.089(0.036)     & 0.107(0.026)    \\ 
20                   & 2000 & 0.028(0.021) & 0.072(0.013)  & 0.169(0.032)  & 0.085(0.064) & 0.141(0.027) & 0.084(0.038)     & 0.101(0.028)    \\ 
30                   & 1000 & 0.033(0.013) & 0.086(0.017)  & 0.198(0.006)  & 0.087(0.056) & 0.108(0.01)  & 0.097(0.039)     & 0.105(0.031)    \\ 
30                   & 2000 & 0.035(0.016) & 0.087(0.016)  & 0.206(0.004)  & 0.099(0.054) & 0.109(0.012) & 0.093(0.037)     & 0.129(0.041)    \\ \bottomrule
\end{tabular}}
\end{table}

\begin{table}
\centering
\caption{Mean(sd) of testing set Bayes error from MINDS and alternative methods, evaluated across 200 replicates. Training sample sizes ranged from 1,000 to 2,000 subjects. The testing sample includes 10,000 subjects. Each model integrated one binary and one continuous modality, incorporating 10–30 binary items and 10 continuous measures.}
\label{table:testing_error}
\resizebox{\textwidth}{!}{

\begin{tabular}{lcllllll}
\toprule
\multicolumn{1}{c}{\multirow{2}{*}{\textbf{Items}}} & \multirow{2}{*}{\textbf{Training Size}} & \multicolumn{6}{c}{\textbf{Testing Set Bayes Error}} \\ \cmidrule{3-8} 
\multicolumn{1}{c}{} &      & MINDS        & Two-step JIVE & K-means      & Hclust       & Two-step K-means & Two-step Hclust \\ \midrule
10                   & 1000 & 0.078(0.041) & 0.158(0.024)  & 0.127(0.042) & 0.147(0.014) & 0.107(0.047)     & 0.139(0.042)    \\ 
10                   & 2000 & 0.067(0.032) & 0.155(0.022)  & 0.119(0.045) & 0.145(0.013) & 0.105(0.049)     & 0.129(0.037)    \\ 
20                   & 1000 & 0.04(0.043)  & 0.15(0.067)   & 0.099(0.067) & 0.116(0.024) & 0.079(0.048)     & 0.094(0.028)    \\ 
20                   & 2000 & 0.038(0.04)  & 0.137(0.067)  & 0.098(0.071) & 0.113(0.023) & 0.079(0.051)     & 0.085(0.022)    \\ 
30                   & 1000 & 0.098(0.03)  & 0.143(0.042)  & 0.097(0.055) & 0.12(0.016)  & 0.088(0.042)     & 0.109(0.021)    \\ 
30                   & 2000 & 0.09(0.019)  & 0.145(0.041)  & 0.093(0.056) & 0.118(0.013) & 0.086(0.044)     & 0.104(0.027)    \\ \bottomrule
\end{tabular}}
\end{table}

The computing time for one iteration of a joint model with 2,000 samples when the number of binary items varies from 100, 200, and 400 items was 0.7, 1.2 and 2.2 seconds, respectively, by Apple M1 Pro.

\section{Disease Subtyping in The ABCD Study}
\label{s:application}
%7. Application to the ABCD Study

We applied MINDS to identify subtypes of subjects in the ABCD study by jointly integrating ADHD- and OCD-related symptoms with neurocognitive behavioral and brain measures. The analysis incorporated multiple modalities: eighteen binary KSADS ADHD symptoms, eleven binary KSADS OCD symptoms, seven continuous brain measures, and seven NIH Toolbox cognition assessments. 
The seven brain measures were the mean cortical thickness values averaged across the right and left hemisphere values. The regions of interest (ROIs) are defined by the
Destriuex atlas \citep{destrieux2010automatic}. The T1-weighted images were preprocessed using
the FreeSurfer 5.1 pipeline \citep{fischl1999high}. We consider seven ROIs which are found relevant to ADHD and OCD diagnosis: the opercular, orbital, and triangular parts of the inferior frontal gyrus; the middle and superior frontal gyri; the long insular gyrus; and the central sulcus of the insula\citep{norman2017shared, brem2014neurobiological, thorsen2020stable}. Among baseline participants aged 9–11 years, greater cortical thickness is considered favorable and reflects healthier neuro-development for each region.
The Toolbox tasks included Picture Vocabulary, Flanker, List Sort, Card Sort, Picture Sequence, Reading, and Pattern Comparison, which capture a range of cognitive domains such as language, reading, working memory, attention, and response inhibition \citep{thompson2019structure}. The higher the Toolbox cognitive scores, the better the performance in neuro-cognitive development. 
In addition, we included three behavioral indices specifically targeting attention and inhibitory control \citep{casey2018adolescent,flankereffect}. These were: (i) the Flanker Cost Effect (mean reaction time for incongruent trials minus that for congruent trials), indexing interference suppression; (ii) the N-back Score (standard deviation of reaction time divided by mean reaction time in the EN-back task), indexing working memory stability; and (iii) the Stop Signal Task (SST) Score (standard deviation of reaction time divided by mean reaction time in the SST), indexing inhibitory control consistency. A detailed description of all cognitive measures is provided in Table~\ref{table:toolbox}.
\begin{table}[!t]
\centering
\small
\caption{Neuro-cognitive measures used in the ABCD study participants subtyping}
\label{table:toolbox}
\begin{tblr}{
  hline{1,2, 12} = {-}{},
}
\textbf{Variable} & \textbf{Task/Test}                               & \textbf{What it measures}                     \\
SST Score                    & The Stop Signal Task                             & Response inhibition                          \\
Nback Score                  & The EN-back Task                                 & Working memory                                 \\
Flanker Cost Effect               & Toolbox Flanker Task                             & Attention, inhibition                          \\
Pic Vocab                    & Toolbox Picture Vocabulary Task                  & Language                                      \\
Flanker                      & Toolbox Flanker Task                             & Attention,  inhibition of automatic response \\
List Sort & Toolbox List Sorting Working Memory Test & Working memory, information processing \\
Card Sort                    & Toolbox Dimensional Change Card Sort Task        & Executive function                           \\
Pattern                      & Toolbox Pattern Comparison Processing Speed Test & Information processing                         \\
Picture                      & Toolbox Picture Sequence Memory Test             & Episodic memory, sequencing                   \\
Reading                      & Toolbox Oral Reading Recognition Task            & Language, reading/decoding skills          
\end{tblr}
\end{table}

There were 11,878 participants enrolled at baseline. We excluded 152 individuals with missing DSM-based ADHD or OCD diagnoses and an additional 548 participants with complete missingness in any of the four data modalities. Participants with full missingness in a modality were removed to avoid potential bias, as such cases may have systematically different distributions. After these exclusions, the final analytic sample comprised 10,126 participants.
Endorsement ratios of individual ADHD and OCD symptoms from the KSADS diagnostic modality is shown in Web Figure 5. ADHD symptoms show greater variability and higher overall endorsement than OCD symptoms. 

\begin{figure}[t]
  \centering
  % Row 1 -------------------------------------------------------
  \begin{subfigure}[t]{0.48\textwidth}
    \centering
    \includegraphics[width=\linewidth]{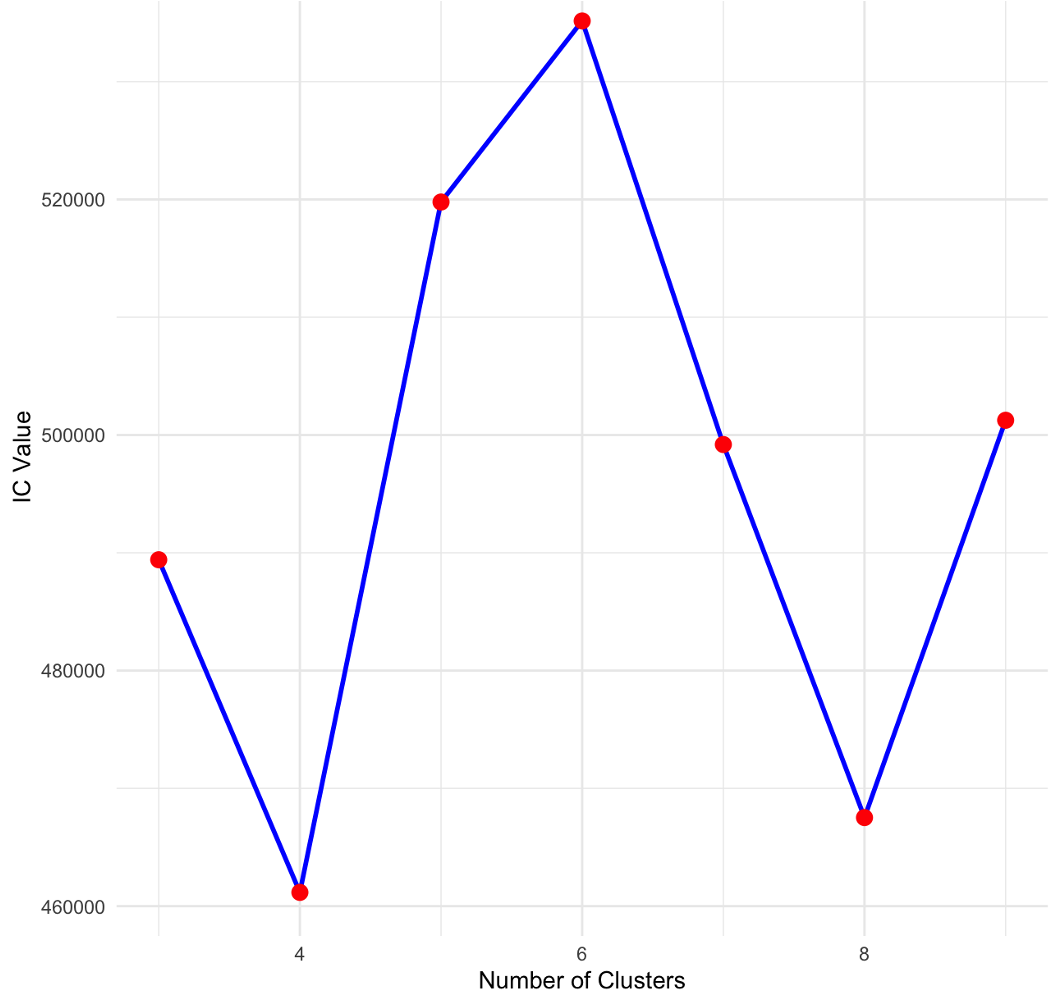}
    \caption{}
    \label{fig:dic}
  \end{subfigure}\hfill
  \begin{subfigure}[t]{0.48\textwidth}
    \centering
    \includegraphics[width=\linewidth]{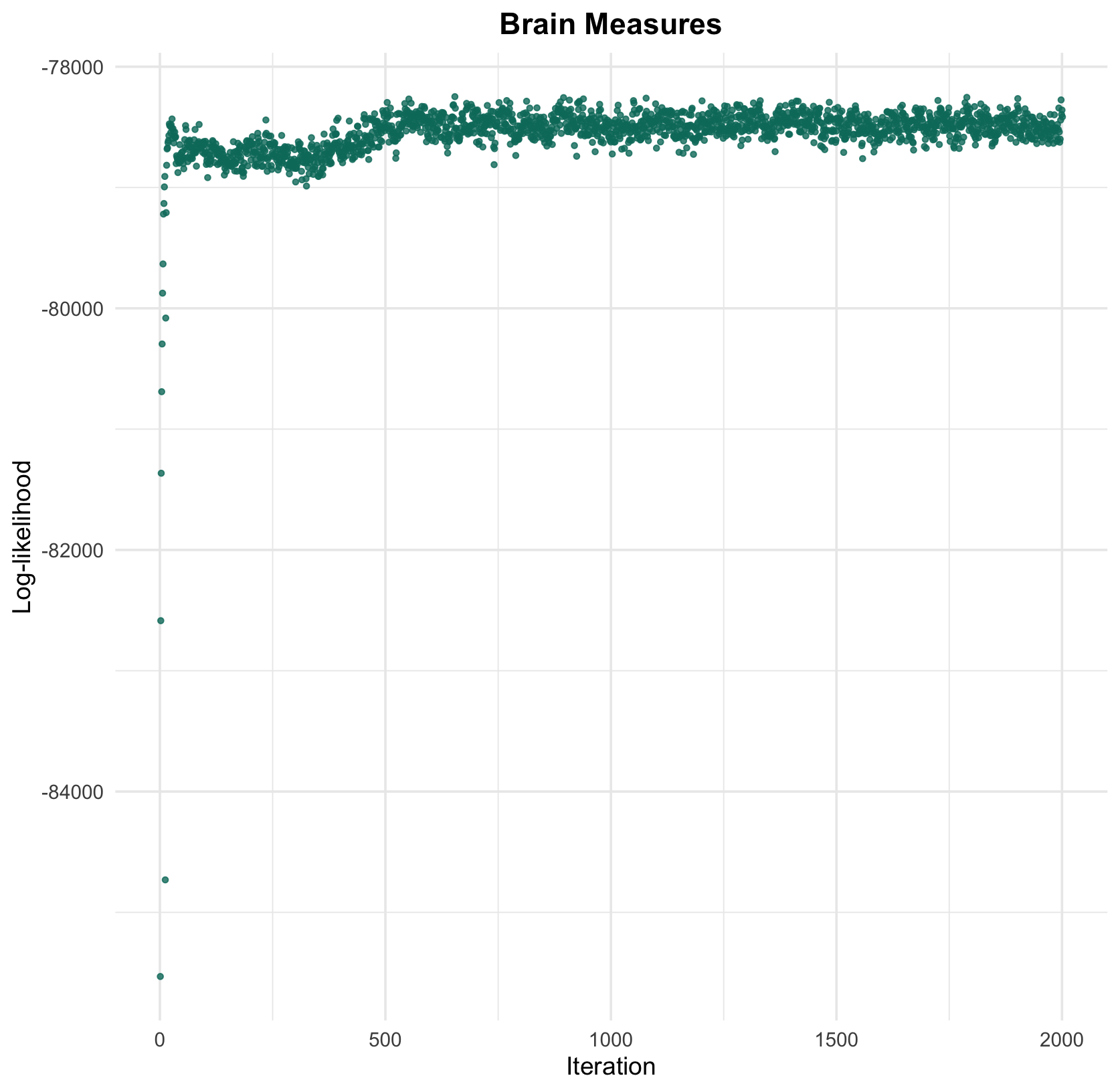}
    \caption{}
    \label{fig:trace1}
  \end{subfigure}

  \vspace{0.6em}
  % Row 2 -------------------------------------------------------
  \begin{subfigure}[t]{0.48\textwidth}
    \centering
    \includegraphics[width=\linewidth]{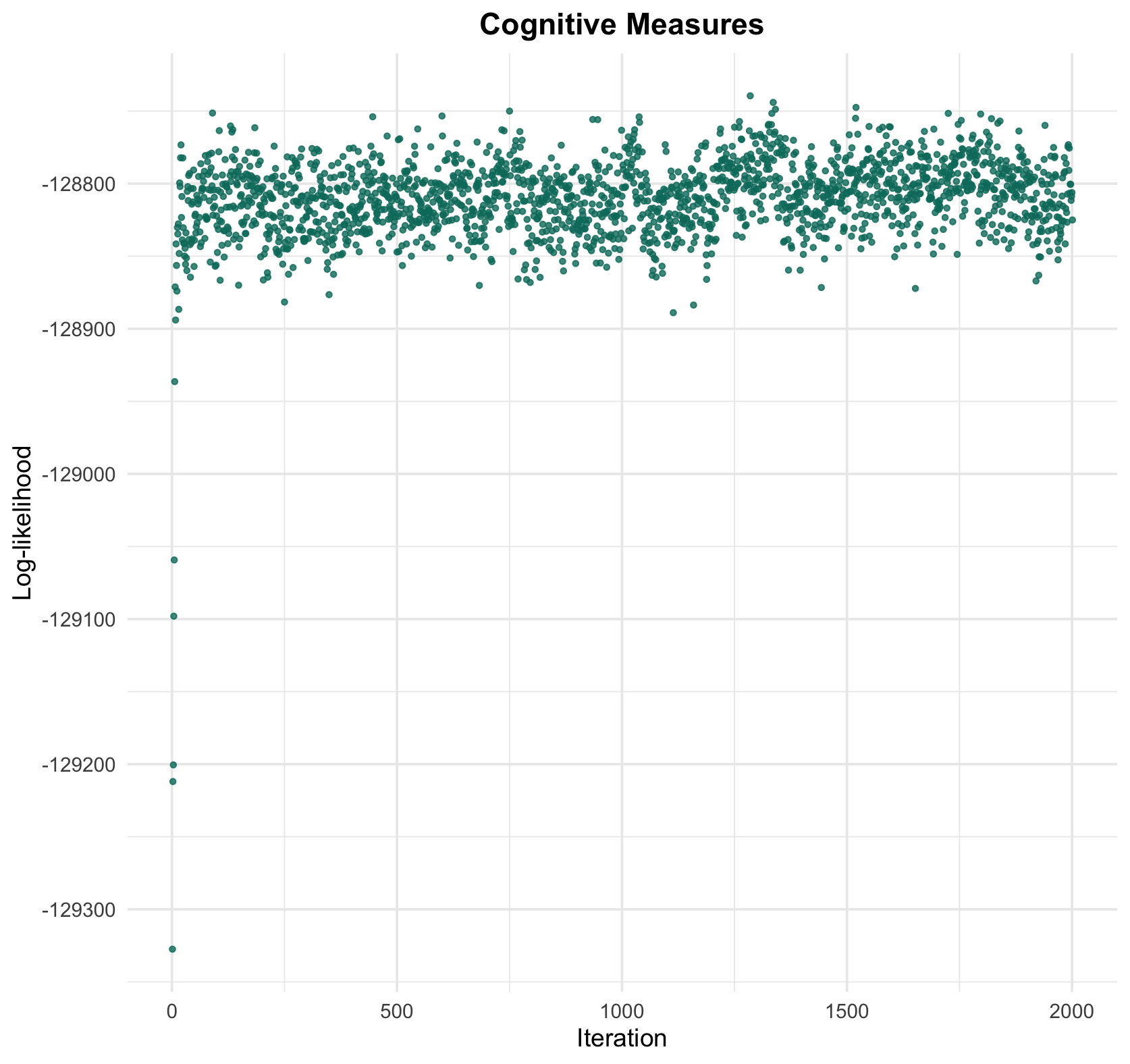}
    \caption{}
    \label{fig:trace2}
  \end{subfigure}\hfill
  \begin{subfigure}[t]{0.48\textwidth}
    \centering
    \includegraphics[width=\linewidth]{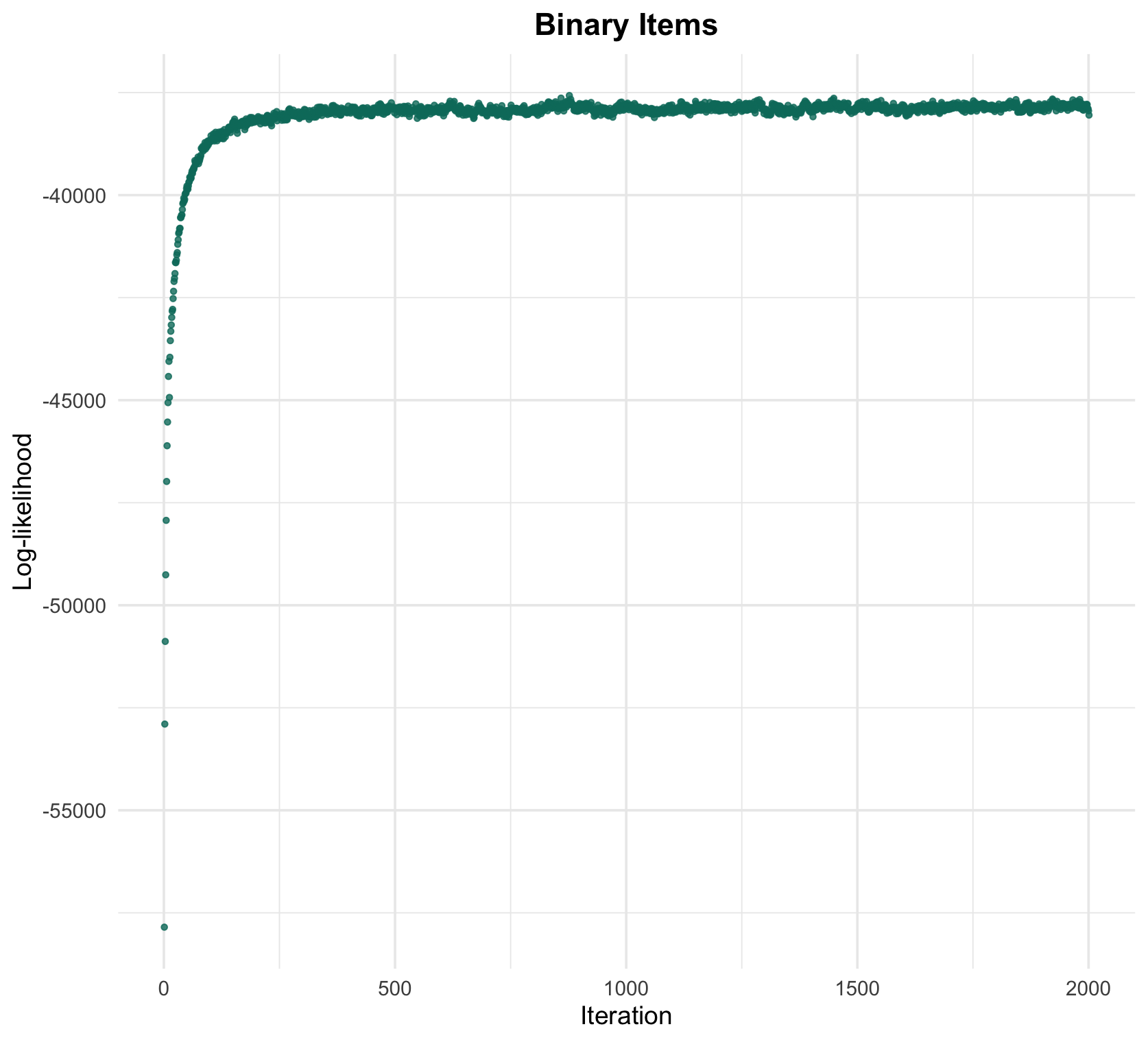}%
    \caption{}
    \label{fig:trace3}
  \end{subfigure}
  \caption{\textbf{Model selection and posterior log-likelihood trace plots across modalities for the real data analysis.}
(A) Information criterion (IC) values are shown for candidate numbers of clusters ($k = 3, 4, 5, 6, 7, 8, 9$), with the optimal model selected at $k = 4$. (B–D) Posterior log-likelihood trace plots for the brain measures, cognitive measures, and binary symptom items, respectively, demonstrate good convergence of the Markov chain Monte Carlo (MCMC) sampler when $k = 4$.}
  \label{fig:model-selection}
\end{figure}

In model fitting, each of the three variability scores was negated so that higher values reflected greater child ability, consistent with the direction of the Toolbox and brain measures. 
We evaluated models with the number of clusters ranging from $k = 3$ to $k = 9$ using the information criterion (IC) to determine the optimal cluster structure. As shown in Figure \ref{fig:model-selection}A, the IC reached its minimum at $k = 4$, indicating that the four-cluster model achieved the best balance between model fit and parsimony. To assess convergence and model stability, we examined posterior log-likelihood trace plots for each modality in Figures \ref{fig:model-selection}B-\ref{fig:model-selection}D. The trace plots for the brain measures, cognitive measures, and binary symptom items all demonstrated stable convergence and consistent mixing of the Markov chain Monte Carlo (MCMC) samples after the burn-in period. These results suggest that the four-cluster solution provides a well-fitted and robust model for integrating multimodal data in the real ABCD analysis.

\begin{figure}[!t]
\centering
\begin{subfigure}{\textwidth}
  \centering
  \includegraphics[width=\linewidth]{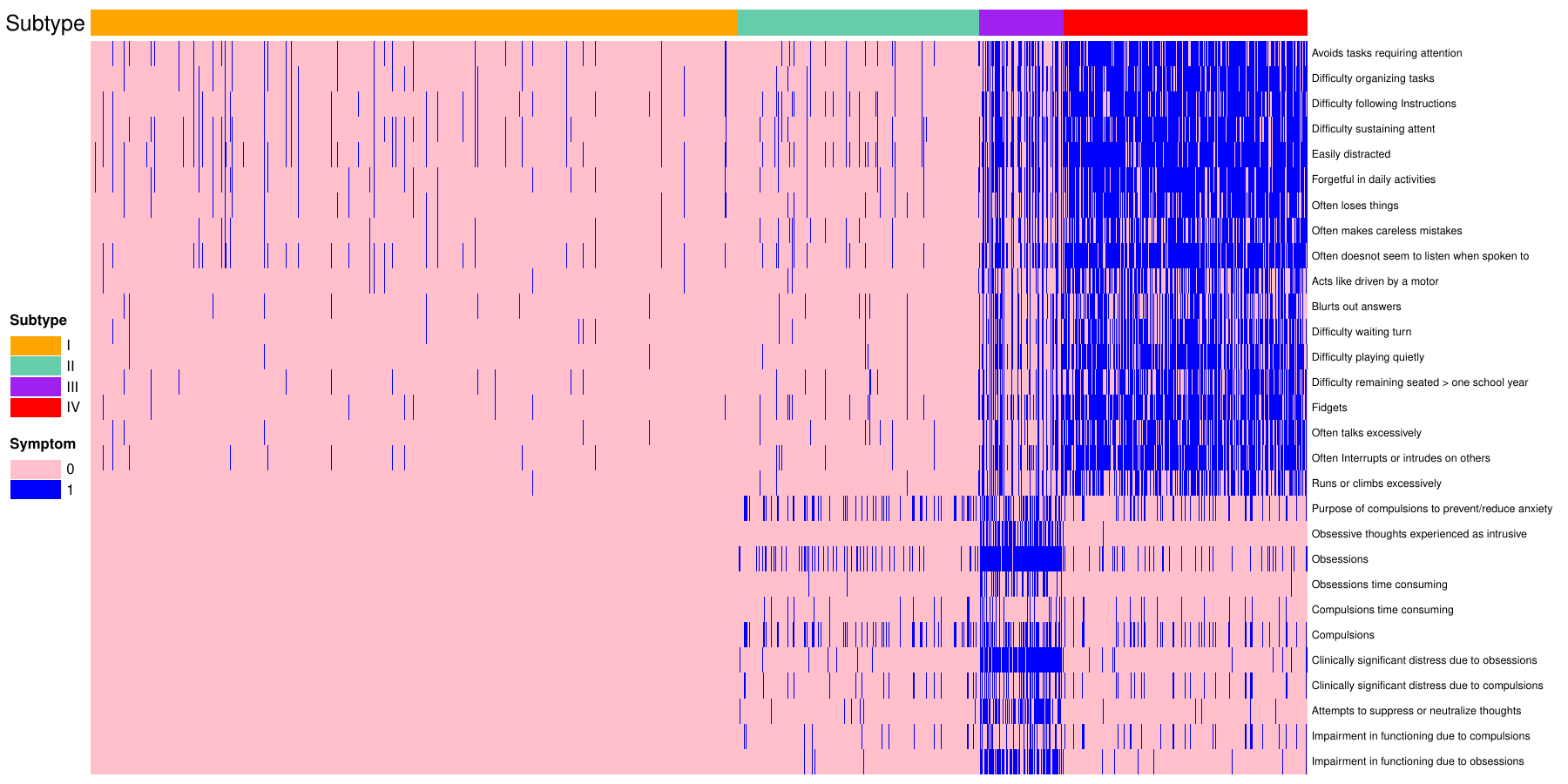}
\end{subfigure}
\begin{subfigure}{\textwidth}
  \centering
  \includegraphics[width=0.9\linewidth]{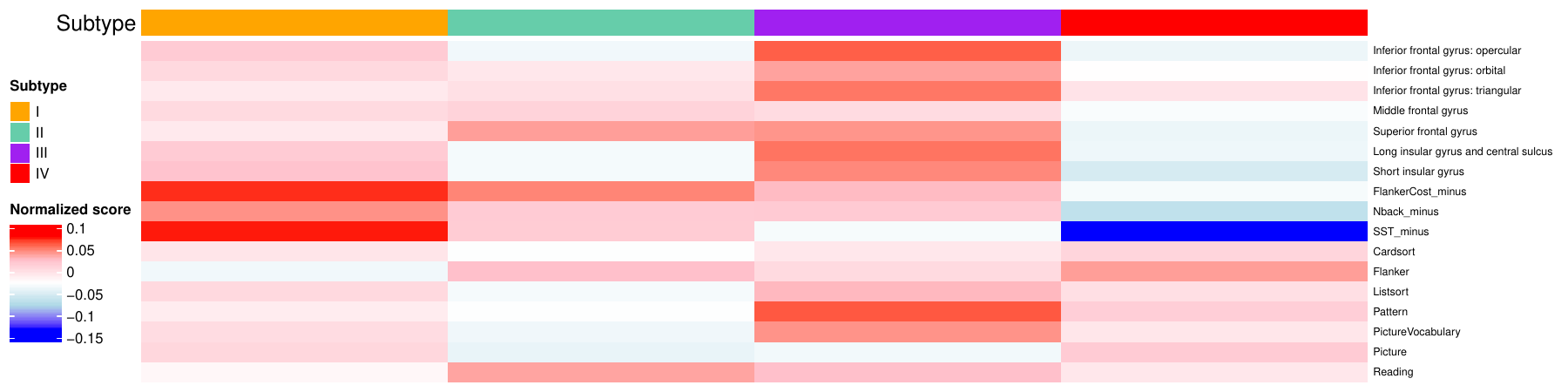}
\end{subfigure}

\caption{Heatmaps of binary items and continuous measures across subtypes by the MINDS method, integrating ADHD- and OCD-related symptoms with selected neuro-cognitive measures. The top heatmap includes eighteen binary KSADS ADHD symptoms and eleven binary KSADS OCD symptoms. Bottom heatmap includes seven cortical thickness measures in selected brain regions, seven NIH Toolbox cognition assessments, and three behavioral indices specifically targeting attention and inhibitory control; scores are normalized by row.}
\label{fig:heatmap_adhd_ocd_behavior}
\end{figure}

Four distinct subtypes were identified by integrating ADHD- and OCD-related symptoms with cortical and cognitive profiles, as illustrated in Figure \ref{fig:heatmap_adhd_ocd_behavior}. Subtype I (Healthy) represented individuals with mild ADHD inattention symptoms and minimal ADHD impulsivity and hyperactivity symptoms. They showed only slightly reduced cortical thickness and relatively intact cognition, with mild deficits in reading and attention, accompanied by efficient attentional control, working memory, and stable inhibition. Subtype II (Mild Symptomatic with Cognitive Difficulties) displayed mild ADHD and OCD symptoms, but substantial cortical thinning across multiple regions. This group demonstrated broad cognitive difficulties across domains, except for Reading and Flanker, while their attentional control and inhibition remained moderately efficient. Subtype III (ADHD\&OCD-dominant) was characterized by severe ADHD symptoms (inattention, impulsivity, and hyperactivity) across all items, alongside pervasive OCD symptoms (obsession and compulsion) and inefficiency in stable inhibition as indexed by SST tasks. Despite this, these individuals exhibited preserved cortical thickness and relatively strong cognitive performance across most NIH Toolbox domains. Finally, Subtype IV (ADHD-dominant with Reduced Brain Development) was marked by severe ADHD symptoms, mild OCD features, and pronounced cortical thinning. Unlike Subtype III, this group showed above-average performance in most cognitive domains, yet exhibited reduced stability of attentional control, working memory, and inhibition.

To compare MINDS with DSM diagnosis, which is the standard criterion for diagnosing ADHD and OCD, we calculated the Calinski-Harabasz Index (CH) as 
$$CH=\frac{B}{W}*\frac{N-K}{K-1},$$ where $B$ is the sum of squares between clusters, $W$ is the sum of squares within clusters, $N$ is the total number of data points, $K$ is the number of clusters. A higher score of CH indicates larger between-cluster variation and smaller within-cluster variation. The MINDS method is better at capturing homogeneity and distinguishing heterogeneity between clusters than clustering by ADHD/OCD DSM diagnosis, which groups the participants into ADHD\&OCD, ADHD, OCD, and None. The Calinski–Harabasz indices (CHs) for continuous measures were 3.45 for the MINDS method versus 1.73 for the ADHD/OCD DSM diagnosis, indicating that MINDS performed 99\% better. For symptom data, the CHs were 3490 for MINDS compared with 1308 for DSM, representing a 166\% improvement.

%%%need update
%Next, we compare MINDS with several other traditional clustering approaches, including K-means, hierarchical clustering (Hclust), two-step K-means, and two-step Hclust. In the two-step K-means, we first conducted factor analysis based on the polychoric correlation of the testing data, and in the second step, we used K-means to cluster the factor scores into $N_c$ clusters. The two-step Hclust differs from the two-step K-means in that the second step uses Hclust to cluster the factor scores.We compared the methods using JCC similarity values (1 - Jaccard distance), where higher values represent greater robustness. The mean (sd) values of the JCC similarity are MINDS: 0.4 (0.05), K-means: 0.19 (0.014), Hclust: 0.3 (0), Two-step K-Means:0.29 (0.016), Two-step Hclust:0.24 (0), based on 100 non-parametric bootstrap samples. Our proposed methods have much higher JCC similarity values, which suggest a greater robustness. 

\begin{figure}
\centering
\includegraphics[width=0.8\textwidth]{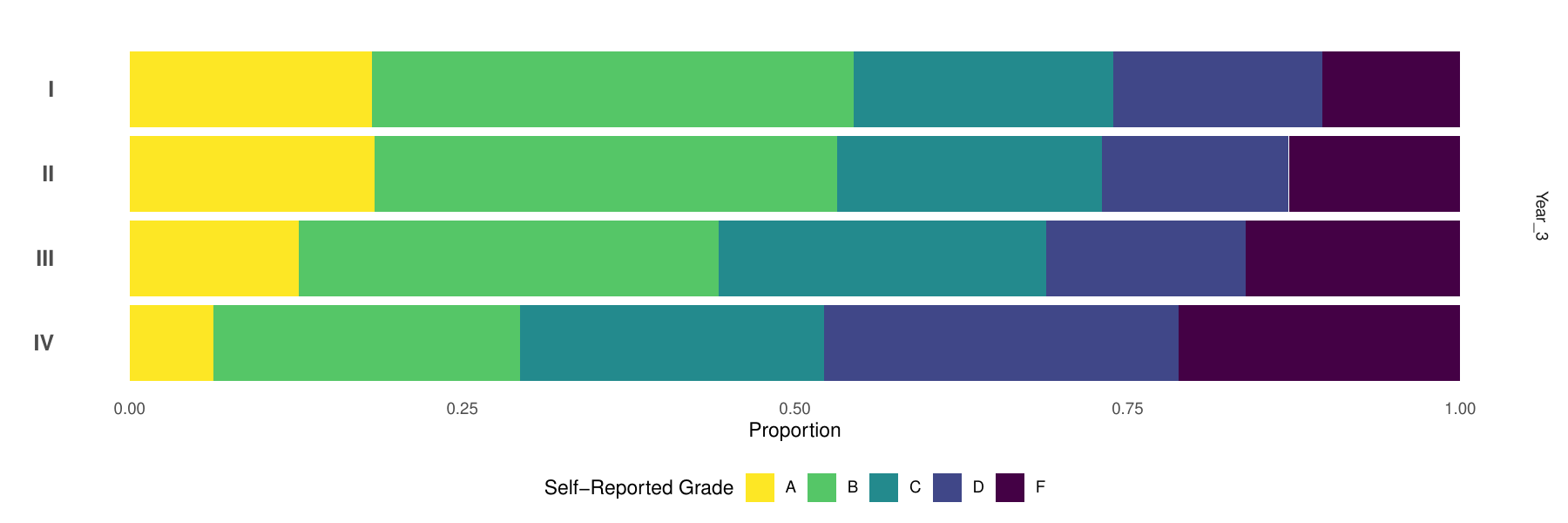}
\caption{Proportion of parent-reported school grades across subtypes at the third follow-up year. Academic performance was assessed using the standard U.S. grading scale, with higher grades indicating better school achievement (A > B > C > D > F), corresponding to numeric scores of 4.0–0.0. This measure serves as an indicator of cognitive and educational functioning in the ABCD cohort. 
}
\label{fig:grades}
\end{figure}

\begin{figure}
\centering
\includegraphics[width=1\textwidth]{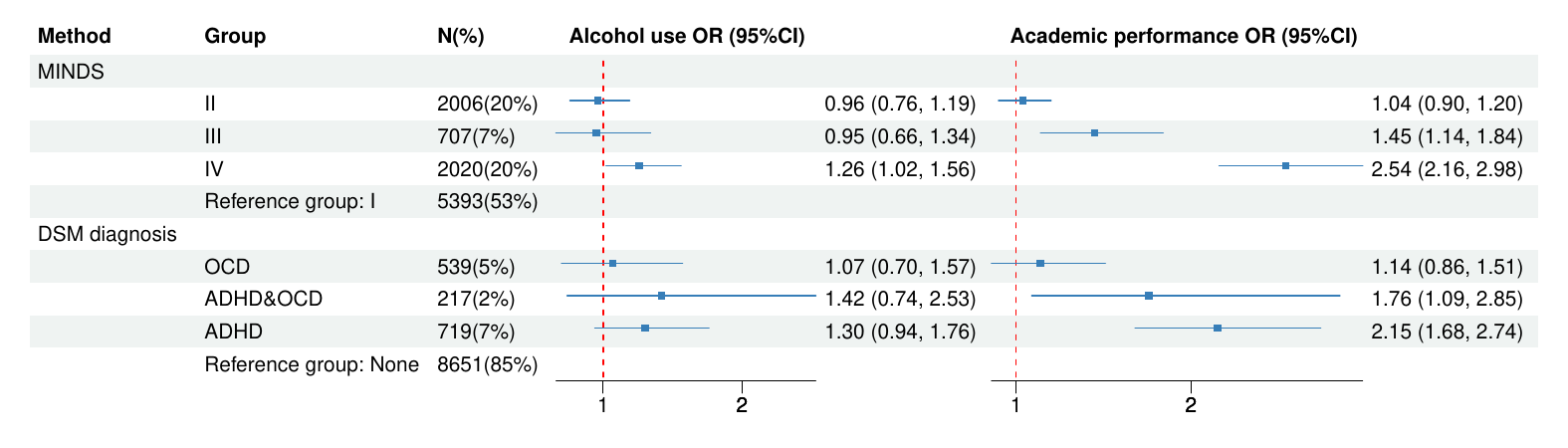}
\caption{
Associations of subtypes with alcohol use and poor academic performance, expressed as odds ratios (95\% CI) versus the healthy group, based on the MINDS method and DSM ADHD\&OCD diagnoses.
}
\label{fig:odds ratio_grade}
\end{figure}

To examine the external validity of the subtypes from MINDS, Figure \ref{fig:grades} shows the distribution of self-reported grades by parents among these subtypes in the third follow-up year. Academic performance was assessed using the standard U.S. grading scale, with higher grades indicating better school achievement (A > B > C > D > F), corresponding to numeric scores of 4.0–0.0. This measure serves as an indicator of cognitive and educational functioning in the ABCD cohort. Our analysis indicates that Subtype I (Healthy) and II (Mild Symptomatic with Cognitive difficulties) exhibit the highest academic achievement. In contrast, Subtype III (ADHD\&OCD-dominant) and IV (ADHD-dominant with Reduced Brain Development) demonstrate the lowest academic performance. 
These results were further confirmed by the results (as shown in Figure \ref{fig:odds ratio_grade}) of the proportional odds logistic regression of the third-year self-reported grades on participants' subtypes adjusted by age, gender, race, education, family income, etc. Subgroup III (OR: 1.45, 95\% CI: 1.14-1.84) and Subgroup IV (OR: 2.54, 95\% CI: 2.16-2.98) are both significantly associated with poor academic performance compared to the Healthy group.

Interestingly, Subtype II, despite exhibiting mild OCD symptoms and broad cognitive difficulties across most Toolbox domains, demonstrated relatively strong academic performance. Notably, it achieved the highest reading score among all subtypes, which may serve as a compensatory factor supporting its overall academic achievement.

The DSM diagnosis subtyping also finds two significant groups, the ADHD\&OCD group (OR: 1.76, 95\%CI: 1.09-2.85), and the ADHD group(OR: 2.15, 95\%CI: 1.68-2.74) associated with academic performance as shown in Figure \ref{fig:odds ratio_grade}. However, MINDS identified 27\% of subjects, compared with only 9\% by DSM diagnosis, as being at risk for poor academic performance and in need of psychological and educational interventions.

Additionally, MINDS results reveal that the Subtype IV has significantly higher odds of alcohol use compared to the Healthy group (OR: 1.26, 95\%CI: 1.02-1.56), shown in Figure \ref{fig:odds ratio_grade}, whereas the DSM subtyping did not find any potentially risky groups. Subtype I exhibits a greater concentration of hyperactivity and impulsivity-related symptoms and less stability in inhibition control, which may explain this association. 

\section{Discussion}

%Disease subtyping is essential for achieving personalized medicine, particularly in mental health, due to the high variability of disorders.
Multimodal clustering analysis is an essential tool for identifying meaningful subgroups beyond psychiatric diagnoses, bridging psychiatric symptoms to biological targets for a better understanding of disease progression and shedding light on future precision treatments.
The ABCD study offers a comprehensive dataset for investigating multimodal integration techniques to understand the heterogeneity of adolescent development. We introduce MINDS, a Bayesian hierarchical joint model with latent variables, to integrate clinical symptoms and neuro-cognitive measures, utilizing Pólya-Gamma augmentation for posterior approximation, which facilitates Gibbs sampling.  MINDS is more advanced than the alternative methods, e.g., iClusterBayes, and other two-step methods, in that it clusters on the original data instead of on intermediate features, which may lead to the loss of shared information, reducing the clustering efficiency. Extensive simulations are performed to assess the consistency of the estimators and the clustering efficiency of our method. We applied MINDS to the ABCD study, where our method demonstrates greater robustness compared to conventional clustering approaches, providing a more reliable classification of clinical subtypes. By identifying meaningful distinctions among ADHD and OCD subgroups, our findings contribute to the development of future psychological and educational interventions tailored to specific clinical presentations in mental health.

There are several challenges to MINDS. 
The estimation of some elements of the loading matrix can have large variances when the initial values are non-informative and the dimensions of binary items and continuous outcomes are high. Using informative initial values, for example, by fitting single-modality models, may reduce variability. As the number of items within each modality increases, computational demands increase, posing challenges in handling high-dimensional data. This can significantly slow down clustering and inference, particularly when there are many variables across modalities. Exploring variational inference methods may mitigate these computational challenges in a Bayesian approach. %Determining the number of clusters may require both data-driven methods (e.g., DIC or IC) and domain knowledge (e.g., clinical interpretability). %Missing data techniques will be considered in simulation and real data analysis in future. 

Lastly, several extensions may be of interest. 
Introducing a penalty to the loading matrix could help manage high-dimensional data more effectively. Additionally, our method could be adapted to integrate other data types, such as zero-inflated or complex-omics data, making it even more applicable to diverse datasets in precision medicine. Validating our identified subtypes in an independent study, such as All{\it of}Us \citep{all2019all} or other clinical samples, would be important.

%%%%%%%%%%%%%%%%%%%%%%%%%%%%%%%%%%%%%%%%%%%%%%
%% Appendix---Please move all appendices to %%
%% a Supplementary file.                    %%
%%%%%%%%%%%%%%%%%%%%%%%%%%%%%%%%%%%%%%%%%%%%%%
%% Support information, if any,             %%
%% should be provided in the                %%
%% Acknowledgements section.                %%
%%%%%%%%%%%%%%%%%%%%%%%%%%%%%%%%%%%%%%%%%%%%%%
\begin{acks}[Acknowledgments]
Data used in the preparation of this manuscript were obtained from the Adolescent Brain Cognitive Development (ABCD) Study (DOI: 10.15154/z563-zd24), held in the National Institute of Mental Health (NIMH) Data Archive (NDA). NDA is a collaborative informatics system created by the National Institutes of Health to provide a national resource to support and accelerate research in mental health. This manuscript reflects the views of the authors and may not reflect the opinions or views of the NIH or of the Submitters submitting original data to NDA.

\end{acks}
%%%%%%%%%%%%%%%%%%%%%%%%%%%%%%%%%%%%%%%%%%%%%%
%% Funding information, if any,             %%
%% should be provided in the                %%
%% funding section.                         %%
%%%%%%%%%%%%%%%%%%%%%%%%%%%%%%%%%%%%%%%%%%%%%%
\begin{funding}
This work was supported by the National Institute of Mental Health [MH123487], National Institute of Neurological Disorders and Stroke [NS073671], and National Institute of Health [TL1TR001875]. 
\end{funding}

%%%%%%%%%%%%%%%%%%%%%%%%%%%%%%%%%%%%%%%%%%%%%%
%% Supplementary Material, including data   %%
%% sets and code, should be provided in     %%
%% {supplement} environment with title      %%
%% and short description. It cannot be      %%
%% available exclusively as external link.  %%
%% All Supplementary Material must be       %%
%% available to the reader on Project       %%
%% Euclid with the published article.       %%
%%%%%%%%%%%%%%%%%%%%%%%%%%%%%%%%%%%%%%%%%%%%%%
\begin{supplement}
\stitle{Supplement A}
\sdescription{Derivation of the Gibbs sampling steps for the joint model.}
\end{supplement}

\begin{supplement}
\stitle{Supplement B}
\sdescription{Supplementary figures and tables for the simulation section and real data analysis.}
\end{supplement}

\begin{supplement}
\stitle{Supplement C}
\sdescription{R code example illustrates the simulation of mixed-type data and the subsequent estimation procedure using the proposed MINDS algorithm.}
\end{supplement}

%%  imsart-nameyear.bst  will be used to                   %%
%%  MR numbers will be added by VTeX.                      %%
%%                                                         %%
%%  Use \cite{...} to cite references in text.             %%
%%                                                         %%
%%%%%%%%%%%%%%%%%%%%%%%%%%%%%%%%%%%%%%%%%%%%%%%%%%%%%%%%%%%%%

%% if your bibliography is in bibtex format, uncomment commands:
\bibliographystyle{imsart-nameyear} % Style BST file
\bibliography{reference.bib}       % Bibliography file (usually '*.bib')

%% or include bibliography directly:
% \begin{thebibliography}{}
% \bibitem[\protect\citeauthoryear{???}{???}]{b1}
% \end{thebibliography}

\end{document}